\documentclass[11pt]{article}
\usepackage[utf8]{inputenc}

\usepackage[margin=1in]{geometry}

\usepackage{setspace}

\usepackage{mathptmx}
\usepackage{bbm,amsmath,amssymb}

\usepackage{graphicx}
\usepackage{caption}
\usepackage{subcaption}

\usepackage{natbib}

\usepackage{comment}

\usepackage{longtable}
\usepackage{multirow}
\usepackage{tabulary}

\usepackage[dvipsnames]{xcolor}
\newcommand{\hl}[1]{{#1}}
\newcommand{\hlpast}[1]{{#1}}

\title{California Exodus? A Network Model of Population Redistribution in the United States\thanks{This work was supported by the National Science Foundation under Grant SES-1826589 and the National Institutes of Health under Grant 1R01GM144964-01.}}

\date{8/11/23}

\newcommand*\samethanks[1][\value{footnote}]{\footnotemark[#1]}

\author{Peng Huang\thanks{Departments of Sociology and Statistics, University of California, Irvine} \and Carter T. Butts\samethanks { } \thanks{Departments of Computer Science and EECS, University of California, Irvine; \texttt{buttsc@uci.edu}}}

\begin{document}

\maketitle

\begin{abstract}
Motivated by debates about California's net migration loss, we employ valued exponential-family random graph models to analyze the inter-county migration flow networks in the United States. We introduce a protocol that visualizes the complex effects of potential underlying mechanisms, and perform \emph{in silico} knockout experiments to quantify their contribution to the California Exodus. We find that racial dynamics contribute to the California Exodus, urbanization ameliorates it, and political climate and housing costs have little impact. Moreover, the severity of the California Exodus depends on how one measures it, and California is not the state with the most substantial population loss. The paper demonstrates how generative statistical models can provide mechanistic insights beyond simple hypothesis-testing.\\


\begin{center} \emph{(To appear, Journal of Mathematical Sociology)} \end{center}
\end{abstract}

%
%

\section{Introduction}
The ``California Exodus'' - a putative phenomenon in which large numbers of individuals are allegedly leaving California and migrating to other U.S. states, has become an increasingly common topic in public discourse surrounding migration and policy in the United States \citep[e.g.][]{bahnsen_great_2021,beam_californias_2021,dorsey_americas_2021,hiltzik_california_2020,song_study_2021}. Popularized within conservative media circles \citep{bahnsen_great_2021,dorsey_americas_2021}, the notion of a ``California Exodus'' serves as a focal point for a political narrative in which the state of California exemplifies the failure of the ruling Democratic party governance, and its associated social and policy regimes. Despite this politicized narrative, the net loss of California population via domestic migration \emph{is} a long-term phenomenon, well-documented in demographic data.  Nor is this a recent development: contrary to popular impression, California's net migration rate has been negative since 1989 \citep{hiltzik_california_2020}. The migration pattern of America's most populous state illuminates important trends of population redistribution in the United States, and could potentially shift the country's economic and political landscape. Historically, internal migration has played a key role in shaping the spatial distribution of population, with the most well-known and general example being urbanization \citep{ravenstein_laws_1885}. In the U.S., internal migration has also played a critical role in its demographic change, including the great migration of African Americans from the South to the North \citep{tolnay_african_2003}, the westward shift of population towards the Pacific coast \citep{plane_migration_1999}, and the ex-urbanization process \citep{plane_migration_2005}. 

Yet, compared to its intense treatment in popular discourse, the California Exodus as a real and persistent (if less dramatic) phenomenon receives scant attention in scientific research \citep[c.f.][]{henrie_Exodus_2008}.
\hlpast{Arguably, this may be in part due to the difficulty of modeling the complexity of internal migration systems, which requires incorporating a wide range of  factors influencing migration. Moreover, as migration systems theory contends \citep{bakewell_relaunching_2014,de_haas_internal_2010,mabogunje_systems_1970}, the migration system has endogenous feedback mechanisms, where migration flows are interdependent to each other. This further complicates mathematical models of migration flows - and their calibration to empirical data - requiring them to account for the autocorrelation structure of the system.} 

In this paper, we use recently developed generative network models of the internal migration system in the U.S. to help unravel the mechanisms sustaining the California Exodus, with an eye to identifying factors that may or may not contribute to this feature of the current U.S. migration system.
We model the U.S. internal migration system as a network comprising counties (nodes/vertices) and migration flows between each directed pairs of counties (edges). Compared to the conventional approach that considers places as analytical units, the relational approach takes migration flows between places as units of analysis, which allows us to capture how the characteristics of origin and destination \textit{jointly} influence their migration flows, such as the difference in political climates and costs of living. The systemic view also considers the endogenous feedback mechanism of the migration system \citep{de_haas_internal_2010}, reflected by the interdependence among migration flows, which gives the system its own momentum, strengthening or ameliorating the exogenous effects from the economic or political landscapes. This is achieved by specifying the network dependence structure, which accounts for the autocorrelation pattern among migration flows. The network models thus can reveal how demographic, economic, political, and geographical characteristics, together with the endogenous feedback mechanisms, shape the direction and magnitude of internal migration flows in the United States.

While computational and statistical constraints have traditionally limited network models of migration to dichotomous or coarsened representations of migration flows, we use recent innovations in valued exponential-family random graph modeling (Valued ERGMs or VERGMs) to estimate a fully quantitative model of interdependent U.S. migration flows at the county level. \hlpast{Motivated by the popular discourse surrounding the California Exodus and existing theoretical and empirical research regarding U.S. internal migration, we focus on four potential social forces that contribute to population redistribution. They include costs of living, political environments, levels of urbanization, and racial demographics.}

\hlpast{This relational view offers new opportunities for insight, but also poses challenges.  For instance, interpretation of the relationship between nodal or dyadic attributes' impacts on migration (i.e., covariate effects) can be complex, as such relationships are subject to both the origin's and the destination's attribute values, and they can take various functional forms. Further, the superposition of forms from multiple effects can make the model difficult to interpret. Such complexities reflect the inherent challenges of capturing an interactive system in quantitative detail, and are thus not unique to migration systems, but are particularly acute when considering networks with valued edges.  We here propose a visualization protocol that showcases how multiple mechanisms involving origin and destination attributes combine to influence the expected number of migrants between origin and destination regions. We utilize this approach to display how the political, racial, rurality, and housing covariates influence the predicted migration flow intensity across different scenarios, offering a quantitative exploration the impact of dyadic factors on migration.}

\hlpast{Another advantage of the VERGM approach is that it offers \emph{generative models,} which can  themselves be used to probe the effects of inferred or hypothetical mechanisms beyond the dyadic level.  Here, we use our empirically-calibrated migration model to perform \emph{in-silico knockout experiments} to investigate how various social, economic, and demographic mechanisms contribute to observed patterns of population redistribution - including, specifically, maintenance of the California Exodus. These knockout experiments simulate migration flow networks under counterfactual scenarios where certain social effects are inoperative \citep{huang_rooted_2022}. Comparing the extent of California's relative net migration loss in the knockout scenarios with that in the observed scenario offers quantitative insights about the impacts of social effects on the pattern of population redistribution.}

\hlpast{The remainder of the paper proceeds as follows.  We begin in Section~\ref{sec:background} with a brief review of different approaches to modeling migration systems, and the extant empirical research that motivates our hypotheses regarding population redistribution in the U.S.} Section~\ref{sec:datamethod} describes the data and variables we use, the model setup including the functional form specification, derivation of the visualization protocol, and the knockout experiment procedure. In Section~\ref{sec:results}, we first offer an overview of the population redistribution pattern in the United States, and the pattern of net migration exchange between U.S. states. We then report our findings regarding the drivers of migration patterns from the ERGM analysis, and show how contributing effects can be visualized. The section concludes with results from knockout experiments. \hlpast{The last section summarizes our empirical findings, our contributions to the mathematical modeling of complex social systems, and some directions for future work.}

\section{Background} \label{sec:background}
\subsection{Modeling Migration Systems}

Migration flows among geographical areas form a complex system, a perspective that has received extensive theoretical discussion in migration studies, in the school of \emph{Migration Systems Theory} \citep[MST,][]{bakewell_beyond_2016,dewaard_resituating_2019,fawcett_networks_1989,kritz_international_1992,mabogunje_systems_1970}. MST introduces two insights regarding migration. First, a migration system consists of flows of people, goods, information, cultures, and other institutions that interact with each other \citep{bakewell_relaunching_2014}. This suggests that understanding migration processes demands a comprehensive survey of various factors and mechanisms, incorporating economic, political, geographical, and demographic analyses. Second, MST emphasizes the \textit{interdependent} feature of migration systems, reflected in their conceptualization of ``internal dynamics'' \citep{de_haas_internal_2010} or ``feedback mechanisms'' \citep{bakewell_relaunching_2014}. The central idea is that there exist endogenous processes, where change in one part of the system can diffuse and alter other parts, creating a systemic momentum. This means that migration flows are correlated to each other. For instance, the migration flow from Seattle to Chicago is associated with the reverse flow from Chicago to Seattle, partly because migrants can carry social connections and useful information from their origin to their destination, motivating and facilitating migration in the reverse direction. Such interdependence among migration flows requires mathematical models of migration to account for the autocorrelation among their observations, and ideally, to also formally and explicitly describe the structure of the dependence.

Researchers have developed various methods to model migration across disciplines including econometrics, geography, statistics, and sociology. A convenient and widely used approach is to treat migration as a feature of areal units, analyzing how the characteristics of a place are associated with marginal migration rates into and out of it \citep[e.g.,][]{partridge_dwindling_2012,treyz_dynamics_1993}. This approach has offered many useful insights and serves as a powerful framework for building predictive models of demographic change \citep{azose_bayesian_2015,azose_estimating_2018}. Methodologically, techniques to account for the autocorrelation in this data structure (areal/lattice data) are well developed in spatial statistics \citep{banerjee_hierarchical_2014}. However, migration is by nature a \emph{relational} process between two places: origin and destination. The above approach by construction marginalizes migration either from an origin perspective or a destination perspective (or condenses both), obscuring how origin and destinations jointly and interactively shape the migration flows between them; such interactions are known to be of considerable importance, as articulated in the classical ``push-pull'' factor model \citep{lee_theory_1966} of migration.  From a network analytic perspective, such models are equivalent to modeling the migration network purely in terms of expected outdegree and indegree effects (sometimes called \emph{expansiveness} and \emph{popularity} in the ERGM literature \citep{holland_exponential_1981}).  Although simple, such models are very constraining - they are essentially similar to a single-dimensional singular value decomposition (SVD) approximation of the adjacency matrix - and are limited in their ability to represent complex structure.

A second model family is the so-called ``gravity model'' (widely used in spatial econometrics), whose unit of analysis is no longer a geographical area but flow within an ordered pair of geographical areas (i.e., an \emph{edge variable}). The original idea of this model family is that the extent of migration flow from origin $i$ to destination $j$ ($M_{ij}$) is positively associated with population sizes in origin and destination ($P_i, P_j$) and negatively associated with the distance between ($D_{ij}$), with the decay usually posited to follow a power law \citep{zipf_p1_1946}, thus superficially resembling gravitational attraction.\footnote{This formulation is also used to describe other types of spatial interactions such as international trade; see e.g. the review of \cite{anderson_gravity_2011}.} Formally, this family is written as 
\begin{equation*}
    M_{ij} \approx C \cdot \frac{P_i^\alpha \cdot P_j^\beta}{D_{ij}^\gamma},
\end{equation*}
\noindent where $C, \alpha, \beta, \gamma$ are positive parameters. Although nonlinear on its original scale, the power law model is intrinsically linear, as shown via the log space representation
\begin{equation*}
    \log M_{ij} = \mu+ \alpha  \log(P_i) + \beta \log(P_j) - \gamma \log(D_{ij})+\epsilon_{ij}.
\end{equation*}
\noindent where $\mu= \log C$ and log error $\epsilon_{ij}$ are unknowns.  Factors other than distance and population size may be incorporated by choosing a suitable regression form for $\mu$. The linear form has facilitated further elaboration, e.g. using a GLM structure to capture discrete outcomes \citep[e.g.,][]{biagi_long_2011}. Although the gravity model does not provide a means of specifying dependence among flows, some extensions in this direction have been proposed \citep[see reviews by][]{patuelli_spatial_2016,poot_gravity_2016}. 

The gravity models have always been in close relationship with network models, with abundant shared knowledge and mutual development.  Fundamentally, gravity models constitute a particular class of network regression models (albeit not necessarily OLS network regression, e.g. \citet{krackhardt_predicting_1988}), a very flexible and successful family.  Substantively, the functional form of the gravity model arises naturally as a model for \emph{tie} (or interaction) \emph{volumes} between regions under power-law spatial interaction functions, a widely observed functional form for interaction probabilities at the individual level \citep{nyerges_spatial_2011}; this, along with the strongly predictive power of distance itself for social networks \citep{butts:ch:2003}, has been argued to provide a mechanistic explanation for why aggregate interactions are often well-approximated by gravity models \citep{almquist_predicting_2015}.  The identification of gravity models with network regression also points to their limitations: while very flexible in specifying relationships between covariates and tie values, network regression models do not specify dependence among edge variables.  While workarounds such as quadratic assignment procedure (QAP) tests \citep{dekker_sensitivity_2007,krackhardt_predicting_1988} can provide statistical answers that are robust to dependence effects, parameterization and/or generation of networks with dependence requires other approaches.

The specification of models for networks with complex dependence among edge variables is a major concern of work on exponential-family random graph models, which we discuss in detail in Section~\ref{sec:ergm}.  ERGMs provide a rich language for specifying interdependencies among edges, as well as associated statistical theory and methodology for inferring such dependencies from observed network data.  Importantly, ERGMs are \emph{generative} - i.e., they provide a full probability model for the target network, and thus can be used for hypothetical realizations of an inferred data generating process.  This makes them especially well-suited to mechanistic investigation using approaches such as \emph{in silico ``knockout'' experiments} and other computational techniques.  The increasing availability of scalable and valued-data ERGMs opens the door to modeling migration systems in a substantively-richer and more statistically-rigorous way.

As noted, one advantage that ERGMs have is the ability to explicitly and formally describe the interdependence of edges within networks. In connection with MST, researchers have utilized this feature to formalize and test the patterns and mechanisms of the endogenous feedback processes in migration systems \citep{huang_rooted_2022,leal_network_2021,windzio_network_2019}. Specifying dependence structure can also improve statistical inference. The autocorrelation among migration flows can not only introduce associations in residuals, but may as well impose more general autoregressive structure. In this case, methods that focus on correcting for correlation in the residuals (e.g., QAP) could be insufficient, running the risk of failing to account for the impact of endogenous factors on covariate effects.

Likewise, the generative aspects of ERGMs are particularly relevant in the context of studying migration systems.  The ability to simulate from empirically calibrated or \emph{a priori} models allows researchers to extrapolate models across spatial and temporal contexts and even investigate counterfactual scenarios. Although there is work in this direction (including applications to the study of migration systems \citep{huang_rooted_2022}), it is arguably an under-appreciated property of this model family, which has been mostly employed as a tool for hypothesis testing. This paper aims to exploit the generative capacity of ERGMs to quantify the contribution of various drivers of population redistribution to the California Exodus.

Despite these advantages, using ERGMs to study migration systems poses a number of challenges. First, it can be computationally intensive to fit (and sometimes to simulate draws from) such models, since closed-form (or even directly computable) expressions for the likelihood are not attainable except in special circumstances. Moreover, generative models for valued/weighted networks are less developed than binary networks, in terms of formal specifications of dependence structures, theoretical justifications of those specifications, and efficient computational tools; this means that researchers sometimes have to dichotomize migration flows, losing critical information about the scale of migration flows. While it is not the focus of the paper to advanced generative models for valued/weighted networks, we employ recent advances in this area to offer a quantitative understanding of the population redistribution pattern within the United States.

Moving beyond ERGMs \emph{per se}, a general challenge in modeling relational data such as migration system data is understanding the combined effects of multiple influences, since prediction of a specific migration flow usually involves attributes from different sources (e.g., origin and destination) that can be combined in different ways. The usual approach of interpreting coefficients separately under the \textit{ceteris paribus} condition is often unhelpful here, as these covariates are intrinsically inter-related.  For example, often it is substantively natural to include covariate factors (e.g., housing costs) of origin, destination, and their absolute difference, where the last term can no longer be interpreted only as a dissimilarity measure since the statistic is fixed once we hold constant the origin and destination covariates. This paper tackles this problem by introducing a visualization protocol that helps interpret the multiplex of inter-correlated functional forms that is common in relational data analysis.

\subsection{Drivers of Population Redistribution}
This section examines possible drivers of population redistribution, with an empirical focus on the case of California Exodus. The first potential driver is the cost of living, suggested by the allegation that people migrate out of California because it is too expensive to live in \citep[e.g.,][]{bahnsen_great_2021,beam_californias_2021}. This is in correspondence to the neoclassical economic theory of migration, that migration happens when the move brings net profit, and lower living costs in destination can be a substantial source of net profit. This motivates our hypothesis:

\textit{H1: The \hl{migration} rate from origins with high costs of living to destinations with low costs of living is higher than the reverse.}

Following the popular narrative that the California Exodus is a political outcome \citep{bahnsen_great_2021}, we hypothesize that political environment could also serve as a driver of population redistribution. Public choice theory and the consumer-voter model consider migration as a means of realizing people's policy preferences \citep{dye_american_1990,tiebout_pure_1956}. Empirical research on U.S. internal migration has also repeatedly observed Americans ``voting with their feet'' \citep{huang_rooted_2022,liu_migration_2019,preuhs_state_1999,tam_cho_voter_2013}. The allegation that Californians leaving their liberal state behind are ``leftugees'' fleeing Democratic governance \citep{dorsey_americas_2021} motivates our second hypothesis:

\textit{H2: The migration rate from liberal-leaning origins (i.e. those with higher share of supporters for the Democratic Party) towards conservative-leaning destinations is higher than the reverse.}

Since population redistribution goes hand in hand with urbanization \citep{lichter_rural_2011,ravenstein_laws_1885}, it is possible that California Exodus is a reflection of the ex-urbanization process. \cite{henrie_Exodus_2008} and \cite{plane_migration_2005} documented the shift of U.S. population from urban areas to rural areas in the 1990s. If this is still happening in 2010s, that might be an underlying mechanism behind California's net migrant loss. We therefore hypothesize that:

\textit{H3: The \hl{migration} rate from urban origins to rural destinations is higher than the reverse.}

Last but not the least, racial dynamics play a critical role in American lives, including migration decisions \citep{crowder_wealth_2006,crowder_neighborhood_2012}.  According to the literature, ``White flight'' is a frequently observed phenomenon \citep{boustan_jue_2023,frey_central_1979,woldoff_white_2011}, where members of the non-Hispanic White population migrate out of racially-diverse places and settle in White-dominant areas. While White flight is associated with the ex-urbanization process, previous literature has identified racial factors to be a unique and non-negligible contributor to this movement \citep{frey_central_1979,kruse_white_2013}. Considering California's diverse racial demographics, White flight could hypothetically contribute to the exodus, and we thus hypothesize that:

\textit{H4: The \hl{migration} rate from origins with low non-Hispanic White concentration to destinations with high non-Hispanic White concentration is higher than the reverse.}

These hypotheses embody a combination of conventional wisdom and notions motivated by migration patterns seen elsewhere.  But are any of them true - and, more importantly, can they account for the California Exodus?  For this, we turn to our empirical analysis.

\section{Materials and Methods} \label{sec:datamethod}
\subsection{Data}
\hlpast{We model the inter-county migration flow network among all 3,142 U.S. counties. The outcome of interest is the average number of migrants moving between each directed pair of counties each year during 2011-2015, which is calculated and released by the American Community Survey (ACS) administered by the U.S. Census Bureau.}

The key covariates capture the characteristics of origin and destination in their costs of living, political climates, level of urbanization, and racial compositions. The cost of living is measured by the median housing costs in 2006-2010 ACS; the political climate is represented by the percentage of voters that voted for the Democratic candidate (Obama) in the 2008 presidential election, as that was the latest national-level election before the study period. The level of urbanization is indicated by the proportion of rural population of a county, estimated by the 2010 Decennial Census. Lastly,  the feature of a county's racial composition is described by its Non-Hispanic White population in the 2010 Census, as this is the most populous racial-ethnic category in the U.S.

The model also considers other covariates that can potentially influence the magnitude of migration flows. The demographic covariates include the (log) population size, log population density (in thousand people per squared kilometers), and age structure (potential support ratio, PSR: ratio of population that are 15-64 years old over population that are 65+ years old), all using 2010 Census Data. The economic covariates include percentage of renters (in contrast to home owners) using 2010 Census, unemployment rates, and percentage of population with higher education attainment, both using 2006-2010 ACS. The geographic covariates include the log distance between origin and destination counties (in kilometers), a dummy variable indicating whether they belong to the same state, and fixed effects for the four major U.S. regions (Northeast, South, Middle West, and West). We also include log migration flow in the previous time period (2006-2010) of the focal migration flow, and the network dependence terms specified in the following section.

\subsection{Valued ERGMs} \label{sec:ergm}
We first model the migration patterns using the valued exponential-family random graph models (valued ERGMs, or VERGMs) \citep{krivitsky_exponential-family_2012}. The ERGM is a parameteric generative model that impose an exponential family distribution to describe the network structure of interest:

\begin{equation} \label{eq:main}
\Pr(Y=y|\theta,X)=\frac{h(y)\exp(\theta^T g(y,X))}{\sum_{y' \in \mathcal{Y}} h(y') \exp(\theta^T g(y',X))},
\end{equation}

\noindent where $Y$ is the random variable of network with realization $y$. $g(\cdot)$ is a vector of sufficient statistics with corresponding parameters $\theta$. The sufficient statistics can be flexibly specified to incorporate both structural covariate effects (e.g., housing price differences) and endogenous dependence terms that capture autocorrelations among migration flows. \hl{In this paper, we include two dependence terms, mutuality and waypoint flow, to account for the endogenous mechanisms that contribute to the symmetry at the dyad-pair level and the node level, beyond the specified covariate effects. Mutuality captures the scale of \emph{reciprocated flow} within dyad pairs ($i\rightarrow j, j \rightarrow i$) by calculating the summation of the minimum edge value across all dyad pairs:
\begin{equation}
  g_m(y)=\sum _{(i,j) \in \mathbb{Y}} \min(y_{ij}, y_{ji} ).  
\end{equation}
\noindent The larger the reciprocated flow within a dyad pair, the larger the statistic. For example, if there are 6 migrant exchange between counties $i,j$, a distribution of \{3,3\} will have the largest reciprocated flow and the corresponding statistic (3), and a distribution \{0,6\} will have the smallest (0). Therefore, a positive coefficient will indicate an endogenous pattern of dyad-level reciprocity, and vice versa. The waypoint flow takes a similar formula, but captures the volumetric flow through each node by examining its total inflows and outflows:
\begin{equation}
    g_f=\sum_{i \in \mathbb{V}} \min \{  \sum_{j \in \mathbb{V}, j \neq i} y_{ij} , \sum_{k \in \mathbb{V}, k \neq i} y_{ki} \}.
\end{equation}
\noindent The larger the volumetric flow moving in and out of a node, the larger the statistic. A positive coefficient will indicate an endogenous pattern of node-level symmetry, and vise versa.}

$h(y)$ is a reference measure that determines the probability distribution of the networks when $\theta \rightarrow 0$. As a VERGM, since our outcome of interest is the count of migrants between two counties, we specify the shape function as a Poissonian reference measure:

\begin{equation}
h(y)=\prod_{(i,j) \in \mathbb{Y}}{(y_{ij}!)}^{-1}
\end{equation}

\noindent This amounts to the assumption that migration events are indistinguishable within edges.  The denominator of the equation~\ref{eq:main} is the normalizing factor that defined on $\mathcal{Y}$, the set of all possible network configurations based on the same vertex set. This intractable function is the source of computational complexity for ERGMs, as it is a function of both the parameter to be estimated, and the set of possible network structures. This is especially the case for VERGMs, since each dyad now can take not only two values for binary networks, but all natural numbers. The more than three-thousand nodes also increases the computational load of our model. To grapple with this challenge, we employ a parallelizable Maximum Pseudo-likelihood Estimation procedure for VERGMs (\citeauthor{huang_parameter_nodate} forthcoming), which is efficient and shows good estimation quality for high-edge-variance networks such as ours. 

\subsection{Functional Form Specification}
There are many possible functional forms for network models even just considering linear formats, since the edge-based models jointly account for the covariates of origin and destination. We thus formulate our key covariate effects based on our theoretical assumptions of their mechanisms that influence migration.

For the cost of living, we include the housing costs of origin and their the difference between destination and origin (destination minus origin). Drawing on the aspiration-ability model of migration \citep{carling_migration_2002,carling_revisiting_2018}, we posit that origin housing costs influences people's financial well-being, which translates into their capacity to migrate; the difference in housing costs influence the utility gain of migrating, altering their aspiration of the migration.

In terms of political, rurality, and racial covariates, we include a dissimilarity measure, implemented as the absolute difference between origin and destination in the corresponding covariate. This follows the operationalization of previous literature \citep{huang_rooted_2022}, which reveals a segmental effect in which less migration happens between counties with larger difference in political climates, levels of urbanization, and racial compositions. Since our interest is population redistribution generated from asymmetric migration, we further include two directional effects. The first is the covariate level of the origin, and the second is a sign function (+1 when destination has a higher covariate level than origin, -1 when the reverse, and 0 when equal).

\subsection{Visualizing Functional Forms}
The composite functional forms of each covariate effect pose the question of how to unpack and interpret their joint effects. We develop a visualization protocol that tackles this problem. For each functional form, the protocol calculates the expected edge value under each possible combination of the covariate value of the origin and destination. To make it more comparable across functional forms, we then normalize it by calculating the ratio of this expected value over the expected value that would be obtained if both origin and destination took the average observed value of the covariate.\footnote{For political, racial and rurality covariates, we use the population-weighted national mean, treating every county as if it had the same share of Democratic voters, non-Hispanic Whites and rural population as the national percentage. For housing prices, we use the national median, as the functional form takes the logarithm of the prices.} We describe this formula as follows. 

In the absence of dependence terms, a Poissonian VERGM is identical to a network regression model with a independent Poisson distributions on each edge \citep{krivitsky_exponential-family_2012}, where there expected value of the $i,j$ edge is:
\begin{equation}
\mathbb{E}(Y_{ij})=\exp(\theta^T \Delta_{ij}^{0\rightarrow1}g(y,X))
\end{equation}
\noindent where $\Delta$ denotes the change in the sufficient statistics when the focal edge's value goes from zero to one. If we only focus on one covariate $f_k(X_{ij})$ (whose sufficient statistic in ERGM will be $f_k(X_{ij})\cdot y_{ij}$), then we have:
\begin{equation}
\mathbb{E}(Y_{ij}) \propto \exp(\theta_k \cdot f_k(X_{ij})) \label{eq:ratio1}
\end{equation}

\noindent so we can express the conditional expected value as a function of origin's and destination's covariate level by calculating the exponentiated product of the functional form and the corresponding coefficient in equation~\ref{eq:ratio1}. We further add a normalizer to center the expected value and make it more comparable across different functional forms. The normalizer is the expected edge value when the covariate of the origin and destination is set to the average value (described in the previous footnote) across the vertex set ($X_0$):
\begin{equation}
\mathbb{E}(Y_{ij})\propto \exp(\theta_k \cdot [f_k(X_{ij})-f_k(X_0)]) \label{eq:ratio2}
\end{equation}
\noindent The formula is in essence the ratio between the expected value of a focal edge under a specific origin-destination covariate vector over the expected value where the origin and destination has the covariate value equal to the average value.

When we need to calculate the ratio for composite expected value, we can simply take the product of their ratios for each form. In the Results section, we will display the functional form of both separate effects (e.g. origin housing costs) and composite effects (e.g. origin housing costs plus difference in housing costs).

Note that this is not exactly the same as the conditional expectation ratio in our specified model, since the model contains dependence terms that distort the edge distribution away from a regular Poisson distribution. A rigorous calculation of the exact expectation ratio is, however, computationally prohibitive, as it requires numerical integration of all possible edge values times their probability function for every realization of the covariate vector. Nevertheless, the knockout experiment in the following subsection takes the dependence into control, offering a closer look at the functioning of the VERGM with dependence terms.

\subsection{Knockout Experiments via Network Simulation}
The visualization of functional forms offers structurally ``local'' insights about how each social force influences migration patterns. Building upon that, we want to quantify how theses social forces contribute to the social phenomenon of interest on a global scale, specifically population redistribution and the California Exodus. We achieve this by leveraging the generative feature of ERGMs to perform \emph{in silico} knockout experiments via network simulation. A \emph{knockout experiment} as employed in a social science context is a model-based thought experiment that examines counterfactual scenarios where certain posited social forces are inoperative, while all other forces are left at their observed levels \citep{huang_rooted_2022}.  The change in outcomes of interest relative to the behavior of the full model is used to probe the impact of the knocked-out mechanism.  Here, we implement knockout of mixing effects by simulating migration flows with all counties having their covariates of interest fixed at an identical value average that is specified in the previous footnote (removing differential mixing). Simulating flows obtained under these conditions, we compare California's ranking in net migration loss across all states under the knockout scenarios with the observed models. This allows us to probe the connection between the mechanisms captured by the model and our social phenomenon of interest. For example, if under the hypothetical condition where every U.S. county has the same housing cost, California's relative net migration loss is not as severe as the observed situation, it would suggest that housing-cost effects on migration could be a contributor to the California Exodus; by turns, if eliminating housing disparities has no impact on asymmetric migration, we can rule it out as a driver of migration loss.

To assist the interpretation of the quantitative results from knockout experiments, we include positive and negative controls in simulation, alongside knockouts of our key covariates of interest: political, racial, rurality, and housing attributes. Originating in the experimental sciences, positive and negative controls are experimental conditions that researchers expect to produce positive and null results, respectively; the controls validate the experimental procedures, serving as the benchmark for other regular experimental settings. In an \emph{in silico} setting, controls remain important to verify that the model is sensitive to manipulations that should have an impact on the outcome of interest (and, by turns, that it is not overly sensitive to manipulations that should not have an impact).  Here, we knock out distance effects as a negative control, treating all dyads as having a common log distance set at national mean. We expect the knockout of non-directional distance effects to not alter the rankings of net migration loss across the country, and the difference between this case and the full model can be considered as a combination of numerical noises and some second-order impacts (since we include complex network dependence terms). The removal of population effects by equally distributing population across all counties serves as a positive control case, as we expect the removal of population effect to have a large impact on the population redistribution pattern. The purpose of these two controls is not substantive interpretation of the fundamental distance and population effects, as the counterfactual scenario is arguably radical and unrealistic, but rather, to provide insights into the question of ``how small is small'' and ``how big is big'' in terms of altering migration ranking.

\section{Results} \label{sec:results}

\subsection{General Patterns of Population Redistribution}

\begin{table}[htbp]
  \centering
  \caption{Annual Population Change in the United States, 2011-2015}
    \begin{tabular}{lrr}
    \hline
    \hline
    & Count & Crude Rate (\%) \\ \hline
    Population & 308,739,316 & \\
    \emph{\textbf{Natural Change}} &       &       \\
    Births & 3,961,037 & 1.28  \\
    Deaths & 2,598,956 & 0.84  \\
    Natural Increase & 1,362,081 & 0.44  \\
    \emph{\textbf{International Migration}} &       &       \\
    Immigration & 1,841,695 & 0.60  \\
    \emph{\textbf{Inter-county Migration}} &       &       \\
    Total migrants & 17,176,675 & 5.56  \\
    Node-level asymmetry & 1,523,550 & 0.49  \\
    Dyad-level asymmetry & 3,844,434 & 1.25 \\
    \hline \hline
    \end{tabular}%
  \label{tab:population_change}%
\end{table}

\begin{figure}[ht!]
\centering
    \begin{subfigure}[b]{.33\textwidth}
        \includegraphics[width=\textwidth]{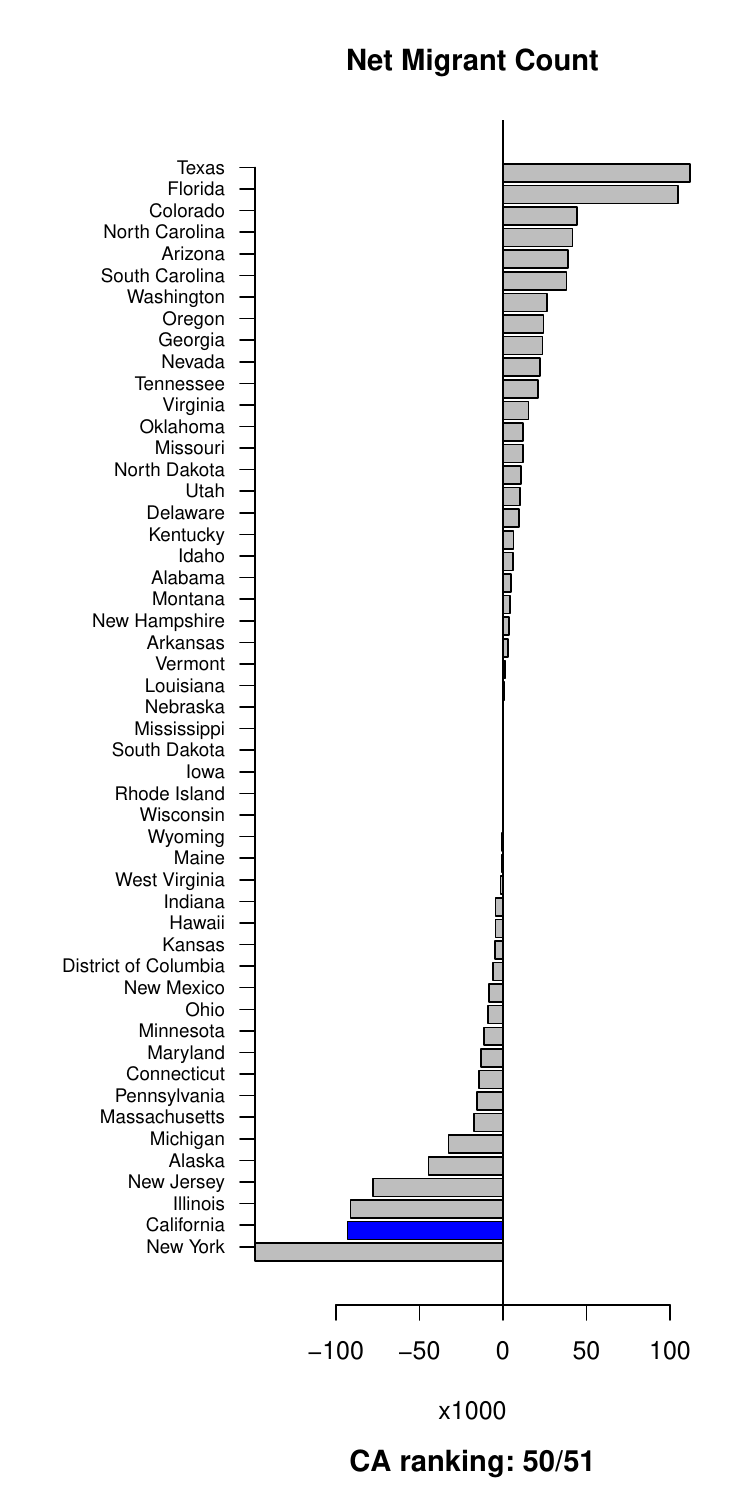}
    \end{subfigure}%
    \begin{subfigure}[b]{.33\textwidth}
        \includegraphics[width=\textwidth]{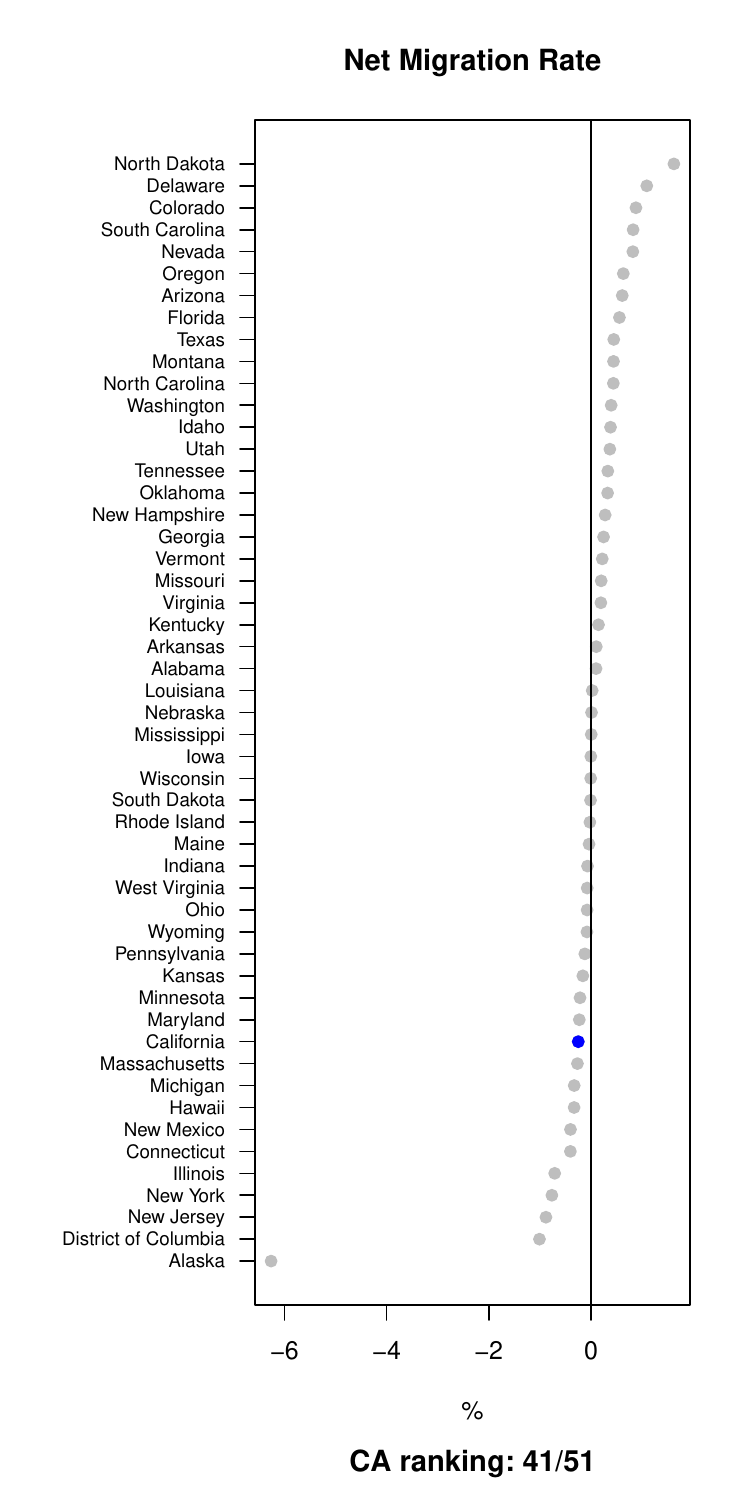}
    \end{subfigure}%
    \begin{subfigure}[b]{.33\textwidth}
        \includegraphics[width=\textwidth]{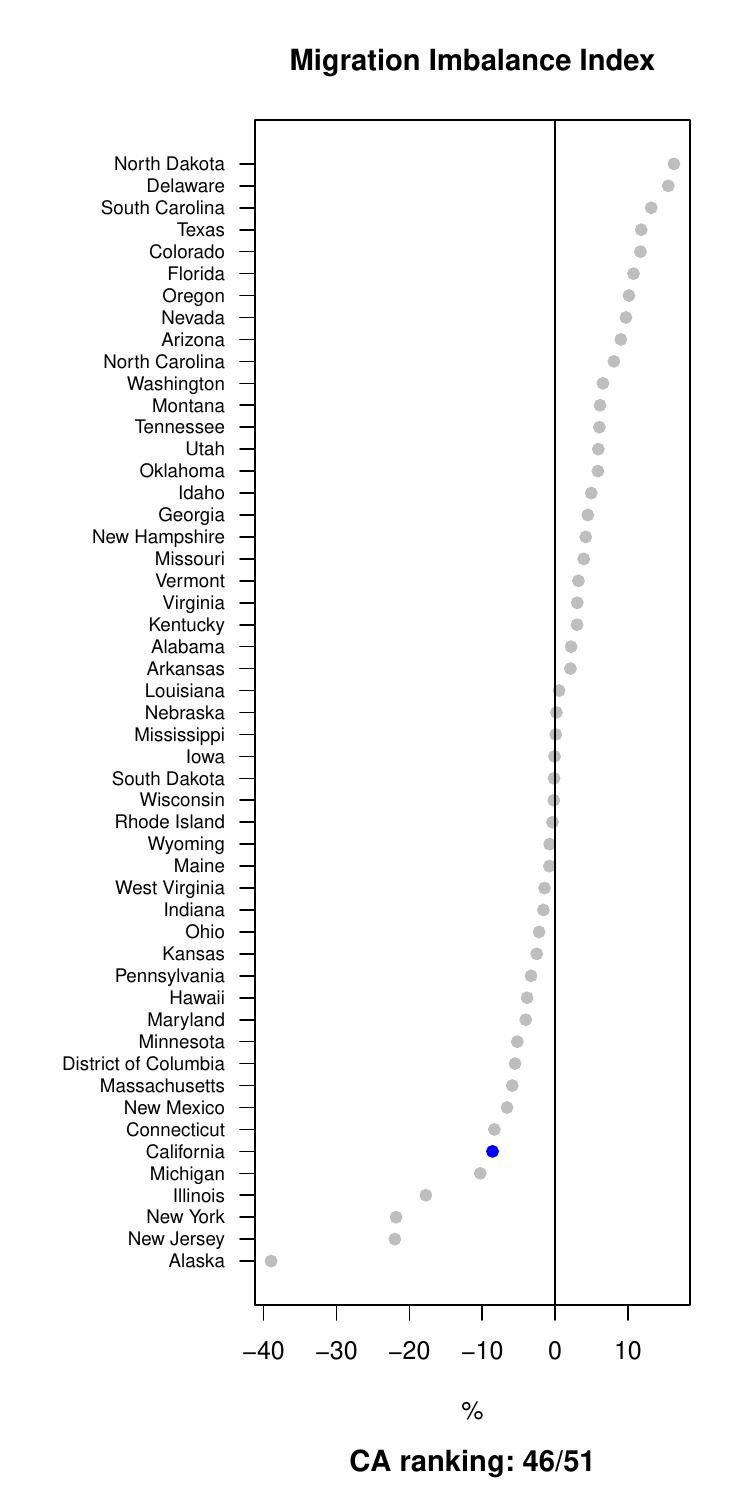}
    \end{subfigure}
    \caption{Net migrant count and net migration rate by state\label{fig:observed_ranking}}
\end{figure}

\begin{figure}
    \centering
    \includegraphics[width=0.5\textwidth]{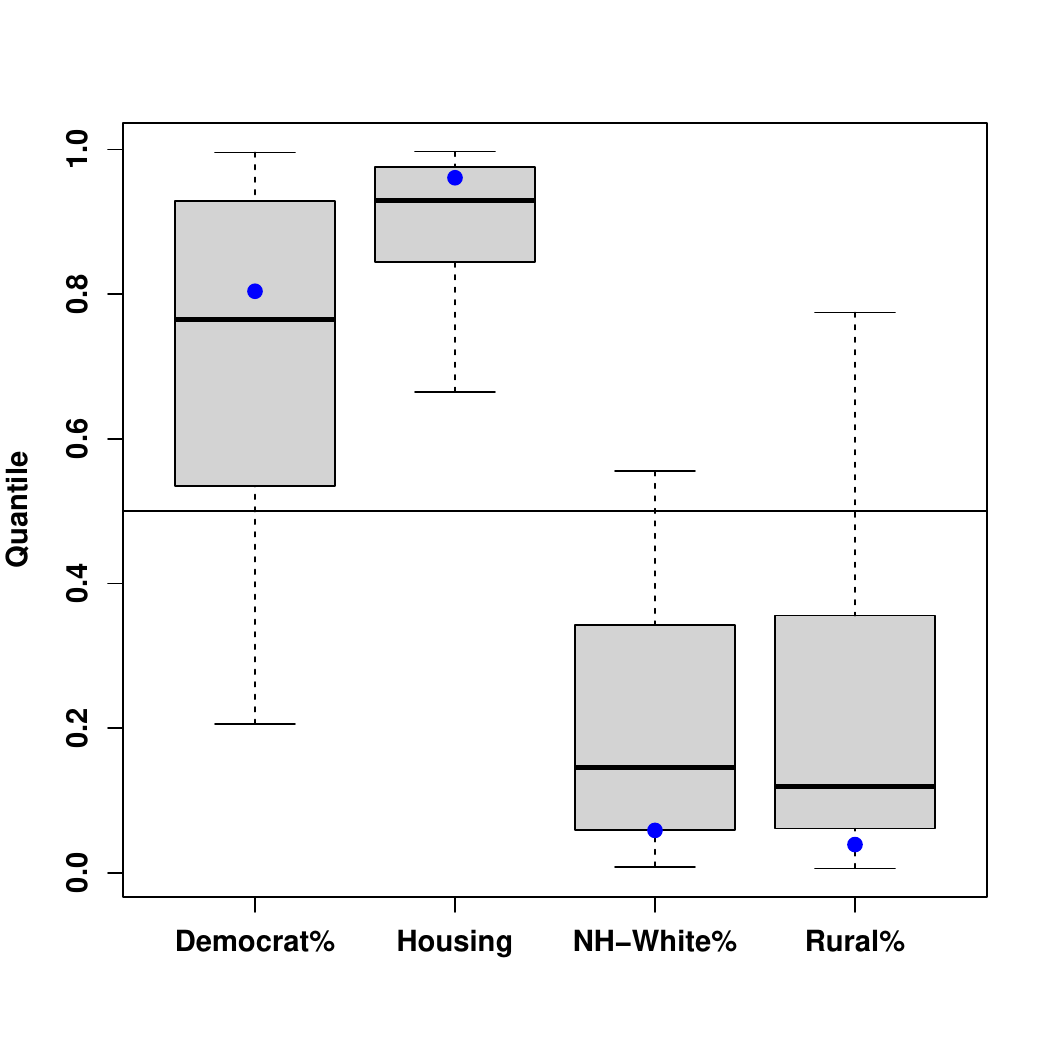}
    \caption{Quantiles of attributes for California counties (boxes) and the state as a whole (blue dots) relative to all U.S. counties}
    \label{fig:quantile}
\end{figure}

To offer a broad view of population change in the study period, Table~\ref{tab:population_change} shows the annual population changes from different demographic processes and their crude rates (normalized by the total population size).\footnote{The population size comes from 2010 Census, the natural change data comes from U.S. Center for Diease Control and Prevention, and the international and internal migration comes from ACS 2011-2015. The natural increase is the number of births minus the number of deaths. The dyad-level asymmetry is the sum of absolute difference across all dyad pairs divided by two: $A_d=\frac{\sum_{i,j}|Y_{ij}-Y_{ji}|}{2}$, and the node-level asymmetry is the sum of absolute difference across all nodes in their inflows and outflows divided by two: $A_n=\frac{\sum_{i}|\sum_i Y_{ij}-\sum_i Y_{ji}|}{2}$. } Compared to natural change and international migration, inter-county migration in the U.S. is a more substantial demographic process with a larger share of population involved. When it comes to population change, the asymmetric internal migration is similar to the scale of immigration and natural increase, all of which have a modest share of population, whcih is around 0.5\% to 1\%. This confirms that as a developed country, the U.S. has a relatively modest population change in the 2010s \citep{rees_impact_2017}.

Figure~\ref{fig:observed_ranking} examines the phenomenon of the California Exodus by comparing the net migrant loss of California (shaded in blue) across three metrics against other U.S. states and the District of Columbia (DC). The left panel displays the net migrant count, which is the total in-migrants minus the total out-migrants. It shows that California has a large net migrant loss, only second to New York among the 51 states and DC. 

Yet, considering the fact that California is the most populous state (roughly 25\% more than the second populous state, Texas, in 2010), the middle panel calculates the net migration rate, which is the net migrant count divided by the state's population. The normalized metric observes California to have a less extreme net migration loss. While it still ranks at the lower end of the list, it is not very different from the majority of the U.S. states, which are within the range of -1\% to 1\%. In other words, the large net outflows of migrants from California can be partly explained by its largest population size. 

Although the middle panel may suggest that there is nothing to be explained - the California Exodus is simply a size effect - examining the \emph{relative asymmetry} of migration to and from California gives a richer picture.  The right panel calculates the migration imbalance index (MII) of each state, which is the net migrant count divided by the sum of in-migrants and out-migrants.\footnote{MII coincides with the migration efficiency/effectiveness index in some migration literature \citep{bell_cross-national_2002,shryock_methods_1973}. It is also directly related to the external-internal (E-I) Index in social network analysis \citep{krackhardt_informal_1988}, although the latter focus on external flows, so MII is equal to one minus the E-I index.} The measurement indicates the proportion of related migrant flows that are inflows of a focal place, capturing the level of imbalance between inflows and outflows of migrants. The right panel reveals that migration imbalance generally has larger variation across states than the net migration rate, as the former focuses on a smaller population, i.e. the migrant population. California has relatively lower ranking in migration imbalance than net migration rate, and its value is farther away from other U.S. states, suggesting a noticeable imbalance in its in/out-migration flows.

In summary, Figure~\ref{fig:observed_ranking} reveals that California is indeed experiencing net migration loss, although the severity relative to other parts of the country vary by the metric we read. Moreover, despite the popularity of the California Exodus narrative, California is actually not the place with the most net migration loss: the New York state has stronger net loss than California across all metrics, and the net migration rate and migration imbalance of Alaska is substantially lower than the rest of the states. These other cases poses important empirical questions that future research should consider.

Lastly, as we consider the possible contributor of California's outstanding net migration loss, we examine California's attributes in Figure~\ref{fig:quantile}. The boxplots shows the quantiles of California counties in those attributes across all U.S. countires, and the blue dots indicates the quantile of California across the 51 states and DC. Compared to other parts of the country, California is indeed a place with stronger left-leaning political environments, expensive housing, larger racial and ethnic minority population share, and higher levels of urbanization. These dimensions are characteristics where California stands out, and therefore has the potential of explaining its migration patterns.

\subsection{Functional Forms of Migration Driving Forces}
\subsubsection{Estimated Effects}

\begin{table}[h]
    \centering
    \caption{Valued ERGM for Inter-County Migration Flows, 2011-2015}
    \label{tab:ergm}
    \begin{tabular}{lr@{\hskip -1pt}lr}
    \hline \hline
    & \multicolumn{2}{c}{Estimate}  & \multicolumn{1}{c}{Std Err} \\ \hline
    \emph{\textbf{Political Covariates}} &&&\\
    Dissimilarity P(Democrat) & -.257 &*** & .007\\
    Origin P(Democrat) & .024 &** & .009 \\
    To higher P(Democrat) & -.008 &*** & .001 \\ 
    \emph{\textbf{Racial Covariates}} &&&\\
    Dissimilarity P(NH-White) & -.172 &*** & .006 \\
    Origin P(NH-White) & -.044 &*** & .007\\
    To higher P(NH-White) & .011 &*** & .001\\
    \emph{\textbf{Rurality Covariates}} &&&\\ 
    Dissimilarity P(rural) & -.457 &*** & .004 \\
    Origin P(rural) & .330 &*** & .006\\
    To higher P(rural) & .018 &*** & .001\\
    \emph{\textbf{Housing Covariates}} &&&\\
    Origin log(costs) & -.283 &*** & .005 \\
    Difference log(costs) & -.148 &*** & .004 \\
    \textbf{Control Covariates} & \multicolumn{3}{c}{(included)}\\ \hline \hline
   \multicolumn{4}{l}{\emph{Note:} **$p<0.01$; ***$p<0.001$ (two-tailed tests).}  
    \end{tabular}
\end{table}

To explain the underlying patterns of intercounty migration, we estimate a VERGM for the migration flow network, with the results of the key covariates of interest listed in Table~\ref{tab:ergm}. The model suggests that, on average, less migration happens between counties with larger differences in their political climates, rurality, and racial compositions as reflected by the percentage of the non-Hispanic White population. In terms of directional effects, the model predicts larger migration flows from counties with higher Democratic Party voter share, and towards counties where the Democratic party voter share is lower. The directionality of the political effects is largely in correspondence to the ``lefugee'' Hypothesis 2 that population are generally leaving from Democratic-party-leaning areas towards Republican-party-leaning areas. The racial effects also run in the direction predicted by the ``White flight'' Hypothesis 4. Holding other factors constant, counties with smaller proportions of non-Hispanic White population send out more migrants, and larger migration flows exist along the way that lead to a county with a higher share of non-Hispanic White population. 

When it comes to rurality, the model is consistent with the ex-urbanization Hypothesis 3 that migration flows are larger when they are moving towards counties with a higher share of rural population than the origin. Yet, the model also shows that counties with higher rurality on average send more migrants out than those with lower rurality. In other words, more migration flows are moving towards more rural regions, but more of them come from a rural county. The housing effects also offer mixed evidence in light of the neoclassical-economic Hypothesis 1. Although migration flows are larger where moving brings greater declines in housing costs from origin to destination, counties with lower housing costs also observe larger out-migration flows. This means that migration typically happens from places with inexpensive housing to places with even less expensive housing.

\hlpast{The model also controls for a series of other covariate effects and endogenous dependence structure, reported in Table~\ref{tab:ergm.full} in the Appendix. The positive mutuality and the negative waypoint flow patterns suggest that, holding other covariate effects constant, the observed migration flow network is more reciprocal at the dyad-pair level and less symmetric at the node level than a random network configuration. This implies the existence of endogenous network patterns discussed in prior literature \citep{leal_network_2021,huang_rooted_2022}. For example, the practice of return migration could promote dyad-level reciprocity, and the signaling effects of county attractiveness can lead to endogenous node-level asymmetry (large migration inflows signaling the popularity of this county, retaining potential migrants from leaving, resulting in an imbalanced in\&out-flow of the county).}

\subsubsection{Visualizing Functional Forms}

\begin{figure}[ht!]
\centering
    \begin{subfigure}[b]{\textwidth}
        \includegraphics[width=\textwidth]{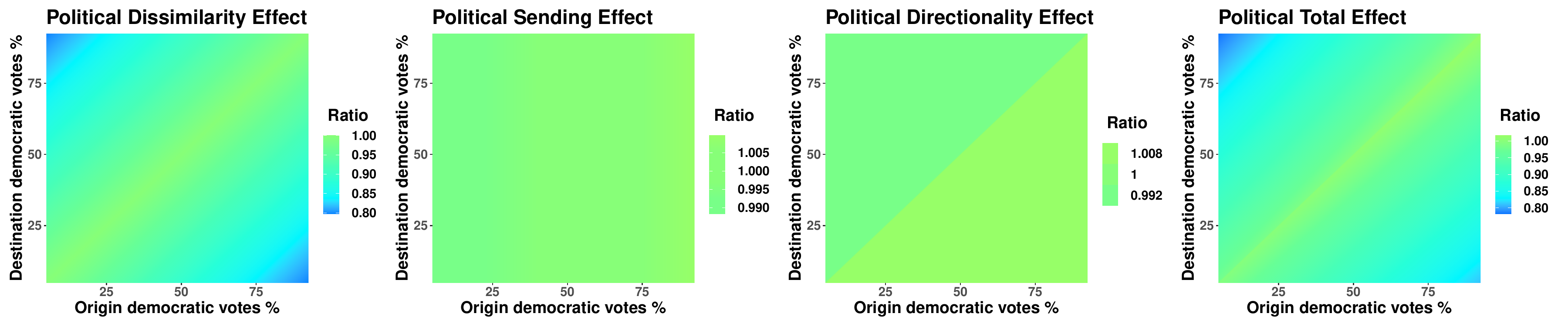}
    \end{subfigure}
    \begin{subfigure}[b]{\textwidth}
        \includegraphics[width=\textwidth]{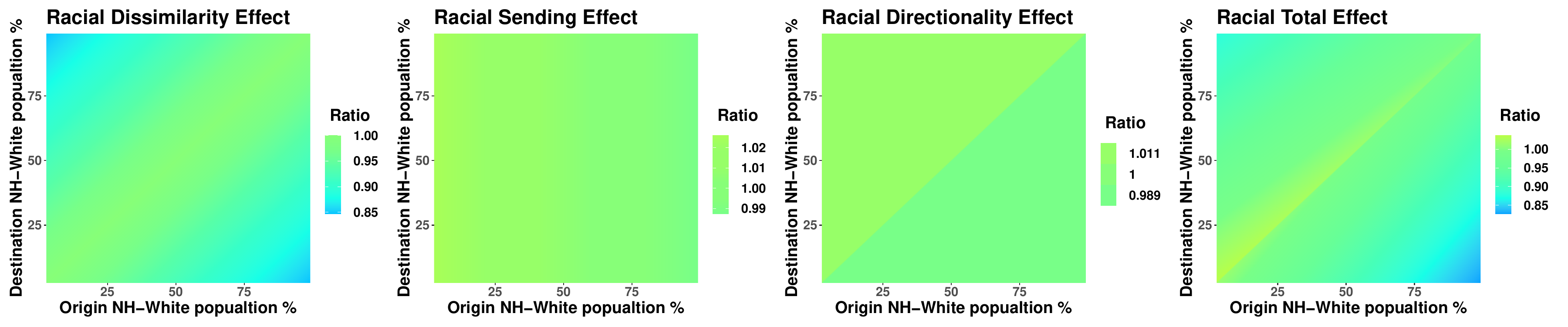}
    \end{subfigure}
    \begin{subfigure}[b]{\textwidth}
        \includegraphics[width=\textwidth]{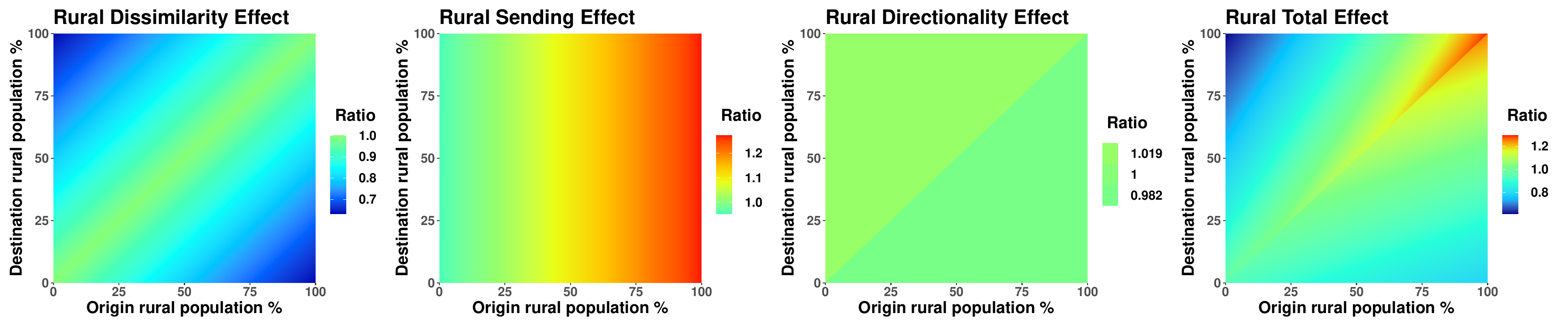}
    \end{subfigure}
    \begin{subfigure}[b]{\textwidth}
        \includegraphics[width=\textwidth]{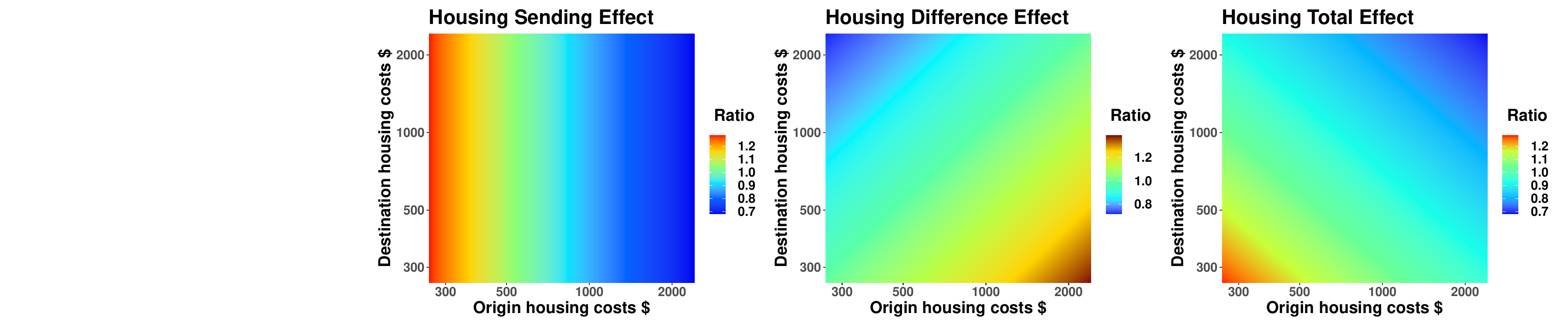}
    \end{subfigure}
    \caption{Function forms for political, rural, racial and housing effects\label{fig:heatmap}}
\end{figure}

For a typical research paper using parametric models, the results section usually stops at the previous subsection, after summarizing whether the directionality of the key effects confirms or refutes the hypothesis. While it is informative to use parametric models as tools for hypothesis testing by evaluating their qualitative behavior, there are more insights one could gain from further the examination of the models. 

First of all, besides the signs of the coefficients and their corresponding $p$-values, their magnitudes also carry critical information about the scale of the effects of interest. Taking the political covariates in Table~\ref{tab:ergm} as an example, the coefficient of origin effects and binary directional effects look an order of magnitude smaller than that of the dissimilarity effect. However, it is difficult to directly interpret the parameter magnitudes, which is subject to the scaling of the covariate distribution.

The second question is about how to interpret holistically the effects of interest, as the different effects (origin, difference, dissimilarity) are interdependent, and holding other factors constant to interpret each single functional form can be unrealistic. This could be a critical question as sometimes different effects offer mixed evidence about substantive hypotheses, such as the rurality and the housing effects in our model. It is of substantive interest to understand how these different effects jointly shape the migration pattern.

To quantify the magnitude of the modeled effects and more concretely understand the separate and joint roles of the functional forms, we visualize the (normalized) predicted migration flow size as a function of origin's and destination's covariate values, displayed in Figure~\ref{fig:heatmap}. Each row presents one chunk of covariate effects, and each column presents a type of functional form, where the higher value in the heatmap indicates the model predicts the migration flow to be higher under these origin-destination covariate values.

The first row of Figure~\ref{fig:heatmap} shows that the directional functional forms (sending and directionality effects) produce very little alternation of the expected migration flow, compared to the undirectional functional forms (dissimilarity effects). The middle two panels show a tiny gradient in its coloring, and the total effects largely resemble the dissimilarity effect, suggesting that the sending and directionality effects make little contribution to the overall effect of political climate. Similarly, in the second row, directional effects of racial covariates also appear negligible, and the undirectional dissimilarity effect dominates the total effect of racial composition. These visualizations tell us that while the directional effects of political and racial covariates run in the direction that correspond to the hypotheses, their effect sizes are small compared to the nondirectional dissimilarity effects.

In the third row of Figure~\ref{fig:heatmap}, although the directionality effect of rurality still resembles those of the previous political and racial covariates, bringing small variation in the expected migration scale, the rural sending effect is strong, and alters the rural total effect to be asymmetric. The bottom row shows that while the sending and difference effects predict substantial variation of expected value across different housing values, their combination offsets each other in the bottom right panel; the gradient of the total effect largely evolves along the $y=x$ line, meaning that swapping the housing costs of origin and destination does not leads to major change in the expected migrant counts. This means that the total housing effect is largely symmetric.

\subsubsection{Visualizing Functional Forms: The San Francisco County Case}

\begin{figure}[p]
\centering
    \begin{subfigure}[b]{.375\textwidth}
        \includegraphics[width=\textwidth]{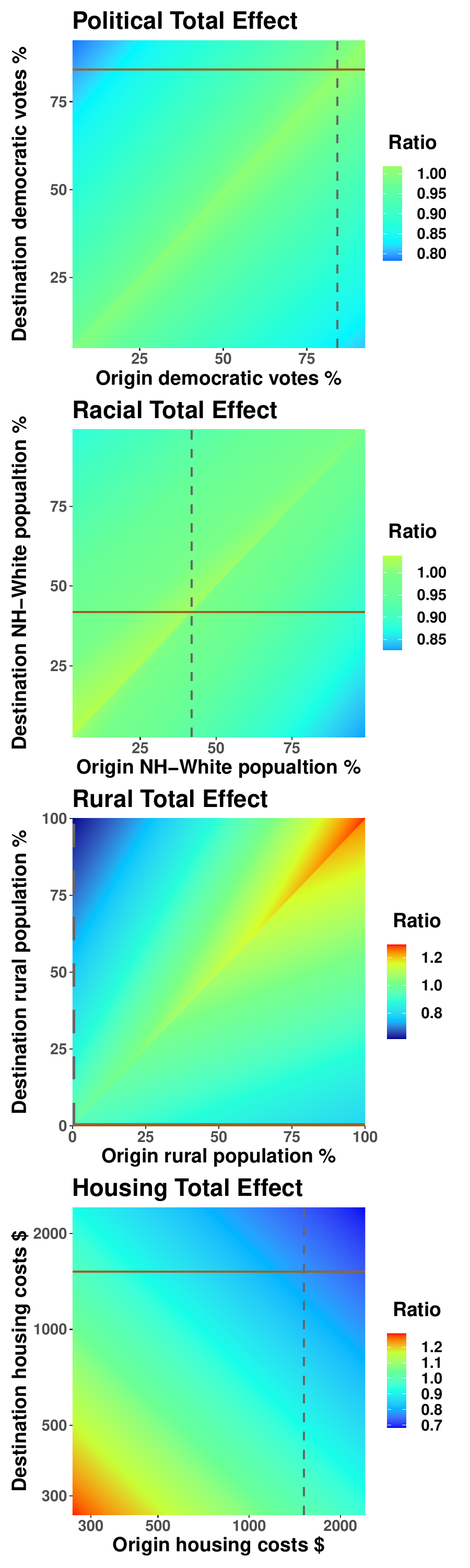}
    \end{subfigure}%
    \begin{subfigure}[b]{.3125\textwidth}
        \includegraphics[width=\textwidth]{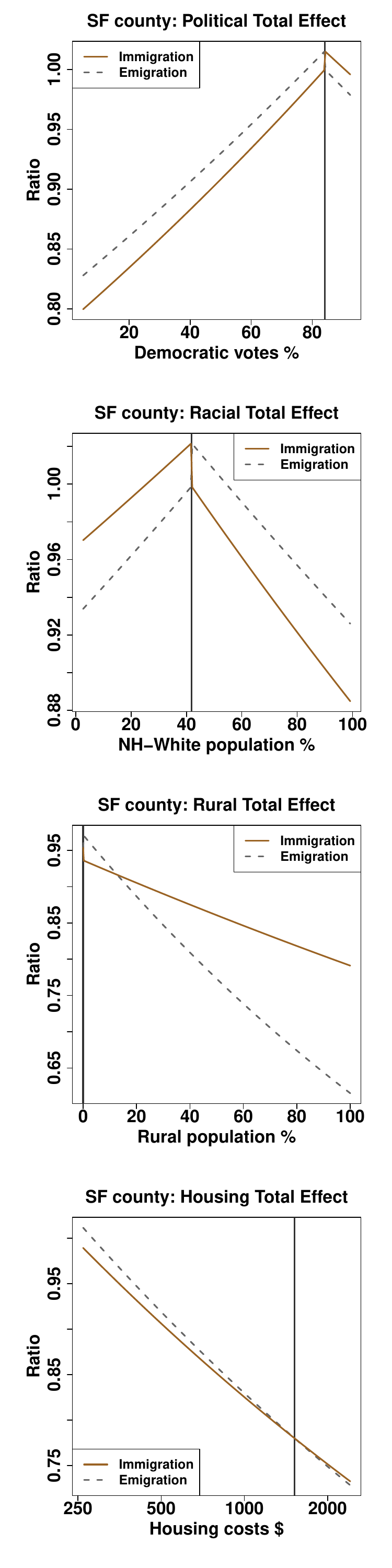}
    \end{subfigure}%
    \begin{subfigure}[b]{.3125\textwidth}
        \includegraphics[width=\textwidth]{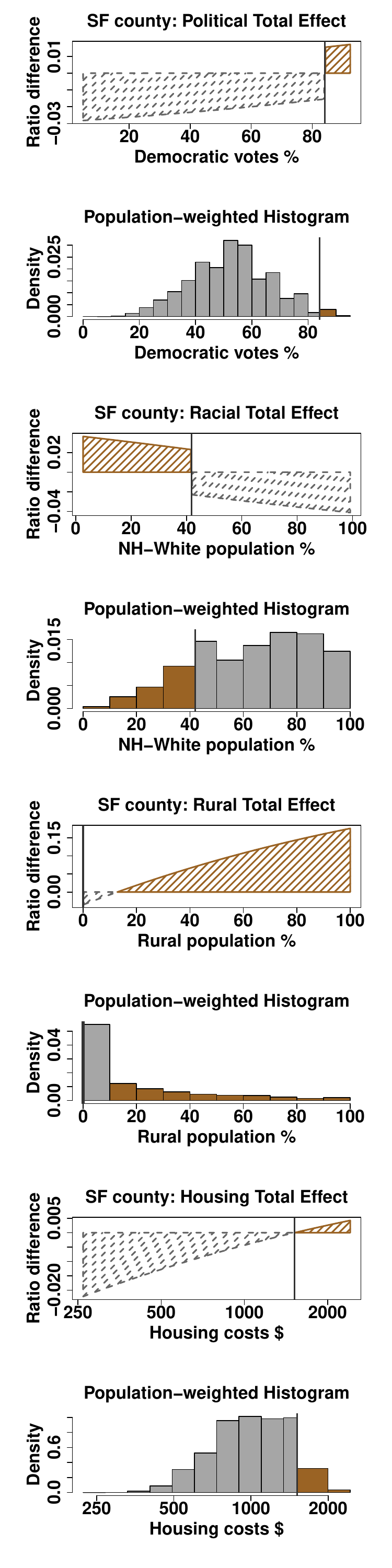}
    \end{subfigure}%
    \caption{Function forms for migration effects involving San Francisco county.  (left) Dyadic effects, with vertical and horizontal lines showing SF attributes.  (center)  Net immigration (solid lines) and emigration (dotted lines) effects for SF, given origin/destination county attributes; vertical line shows SF position.  (right) Areas between curves (net immigration) from the center plot by origin/destination county attributes; histograms show population-weighted distributions of U.S. counties, with brown columns indicating population in net SF-immigration counties.  \label{fig:SFfunction}}
\end{figure}

To further aid our interpretation of the total effects, Figure~\ref{fig:SFfunction} examines the case of San Francisco (SF) county, California, and evaluates its expected migration flows towards and from other counties based on their corresponding covariate value. The first column is a replication of the last column in the previous figure, but adds reference lines that indicate the covariate level of SF county. The middle column extracts from these two reference lines and plot the expected number of immigrants to (brown solid lines) and emigrants from (grey dotted lines) SF county as a function of the origin/destination county's covariate level. The upper right panel of each row summarizes the middle column by getting the difference of immigrant ratio and the emigrant ratio, where a positive ratio difference (shaded in solid brown lines) suggests an expected net migration gain for SF county, while a negative ratio difference (shaded in dotted grey lines) suggests an expected net migration loss for SF county. The bottom right panel of each row plots the histogram of U.S. population about the covariate level of their residing counties. The juxtaposition of the last two plots reflects whether the country's population gravitate towards counties that SF county has net migration gain from (shaded in brown), or counties that SF county has net migration loss towards (shaded in grey), offering a first-order approximation to whether the social effects promote or suppress population loss from a county like San Francisco.

Focusing on the right column of Figure~\ref{fig:SFfunction}, we observe that SF county receives net migration gains from counties with more Democratic-party voter share, which comprise a small share of U.S. population. By turns, it loses migrants to counties with less Democratic-party voter share, which comprise a large share of U.S. population. 
Similarly, in the second row, SF county receives net migration gains from counties with less non-Hispanic White population share, which comprise a small share of U.S. population. The functional form of rurality for SF county is a bit more complicated, as the county takes the extreme value of 0\% rural population. The county is expected to have no net migration exchange with other counties that have 0\% rural population, which consist 7\% of the total U.S. population. SF county is expected to lose population to counties with rural population larger than zero but smaller than 13\%, which includes about 51\% of the total U.S. population. In other words, on average, there are slightly more persons residing in counties that SF county has net migration loss towards. However, once the county deviates from the extreme case of the fully urbanized, the trend reverses, with more of the U.S. population residing in places from which the focal county has net migration gain. Lastly, the bottom right panel shows that the majority of the U.S. population resides in counties with cheaper housing than SF county, areas to which SF would be expected (\emph{ceteris paribus}) to lose population.  Overall, for the SF county case, across all covariates, the model predicts an overall net migration loss from SF county; this is not because all factors \emph{unilaterally} favor emigration from SF, but rather because in each case SF's attributes favor immigration from a relatively small number of counties (with relatively low total population) relative to those to which they favor emigration.

\subsection{Knockout Experiments for the California Exodus}

\begin{table}[ht]
\centering
\caption{California's Average Simulated Ranking with and without Knockouts, by Metric} 
\begin{tabular}{lcccccc}
\hline \hline
& \multicolumn{2}{c}{Net Migrant Count} & \multicolumn{2}{c}{Net Migration Rate} & \multicolumn{2}{c}{Migration Imbalance} \\
  \cline{2-7}
 & Ranking & Change & Ranking & Change & Ranking & Change \\ 
  \hline
Full Model & 50.00 & & 42.08 &  & 45.55 &  \\ 
  \textit{Remove Distance Effect} & 50.00 & 0.00 & 41.92 & -0.16 & 45.35 & -0.20 \\ 
  \textit{Remove Population Effects} & 48.75 & -1.25~ & 28.30 & -13.78~~ & 46.63 & +1.08 \\ 
  Remove Political Effects & 50.00 & 0.00 & 42.00 & -0.08 & 45.16 & -0.39 \\ 
  Remove Racial Effects & 50.00 & 0.00 & 39.36 & -2.72 & 42.49 & -3.06 \\ 
  Remove Rurality Effects & 50.00 & 0.00 & 43.00 & +0.92 & 47.49 & +1.94 \\ 
  Remove Housing Effects & 50.00 & 0.00 & 41.92 & -0.16 & 44.95 & -0.60 \\ 
  
   \hline \hline
\end{tabular}
\label{tab:CAknockout}
\end{table}

The visualization of covariate effects offers us some quantitative insights about how different social forces operate across different origin/destination pairs.  However, our examination of the SF case underscores the intuition that the way in which such forces play out depends upon the global distribution of population (and covariates), which is challenging to infer from direct inspection. For instance, the high level of urbanization in SF county makes it an interesting but special case, and it becomes difficult to visualize every possible rurality level that California counties take and integrate them to offer a holistic evaluation of the rurality effect on the California Exodus. Building on these exploratory insights, this section aims to explicitly examine the connection between migration patterns incorporated into the model with specific social outcomes of interest, such as the California Exodus.

We achieve this by performing \emph{in silico} knockout experiments, with results displayed in Table~\ref{tab:CAknockout}. The first column suggests that California's ranking in net migrant count stays constant throughout all the knockout scenarios except the positive control that knocks out population, contributing to a 1.25 position improvement its ranking (smaller ranking means less net migrant loss). Notice that only knocking out population effects in the positive control alters California's average ranking in net migrant count, and that in the second column, the net migration rates under normalized state population lead to fluctuations of California's average rankings under all knockout scenarios. This suggests that California's status as the largest U.S. state is a major explanation for its substantial net emigration in absolute terms.

In the second column of Table~\ref{tab:CAknockout}, the removal of political and housing effects improves California's ranking in net migration rate at a scale smaller than or roughly equal to the negative control of removing distance effects. Although political and housing effects seem to operate in a direction that contributes to California exodus as hypothesized, their influence on net migration rate is substantively negligible. Knocking out racial effects and rural effects improves and worsens California's relative net migration rate, respectively, indicating that racial effects contribute to California Exodus (from a migration rate angle), while rural effects actually buffer California from even larger population loss. These two changes are larger in their scale than the negative control of distance effects, but not comparable to the positive control of population effects, suggesting their impacts to be moderate.

The last column in Table~\ref{tab:CAknockout} shows California's ranking of migration imbalance. As with the case of net migration rate, removing political, housing, and racial effects reduces California's relative migration imbalance, while removing rurality effects worsens it. Quantitatively speaking, the impact of knockouts of political and housing effects are again similar to that of the negative control of distance effect, while the removal of racial and rural effects bring a ranking change even larger than that from the positive control case of population effects. The small alteration from the positive case is understandable, as the origin and destination effects of population are not hugely different in our model (as well as in many other gravity models, \cite{boyle_exploring_2014}); while changing the total size of migrant population (symmetrically) can alter state rankings of net migration rate given a constant total population denominator, for migration imbalance that solely focuses on the migrant population, this is no longer the case. The fact that none of the knockouts alters California's relative migration imbalance in a sizable way suggests that California's migration imbalance does not result from one single social effect.

\section{Discussion}
Leveraging a large-scale valued network model, this paper studies population redistribution patterns in the United States, and in particular the heatedly discussed case of the ``California Exodus.'' Our analyses show that California indeed experienced net migration loss in the 2010s, although its scale varies depending on the metrics one examines; the exodus is substantial in absolute terms but relatively small in its crude rate (count per capita), while still being fairly considerable in its imbalance between in-migration and out-migration flows. Valued ERGM  analysis reveals the direction of the political, rural, racial, and housing effects on population redistribution, which largely work in directions that would contribute to net migration loss for highly populous counties like San Francisco. Knockout experiments further show that racial effects contribute to the California Exodus, rurality effects work \emph{against} the California Exodus, and while political and housing effects contribute to the California Exodus, their effects are largely negligible. The scale of these effects on the California Exodus varies by the migration metric used, but none of the knockout scenarios (except a positive control case for population distribution) alter California's ranking in net migration loss in a substantial way. This suggests that the California Exodus is not governed by one single social effect, but is a joint outcome of complex systemic patterns.

Methodologically, this paper offers a roadmap that aids interpretation of composite functional forms in parametric relational models via visualization. It also demonstrates the insights generative models such as ERGMs could offer by designing simulation experiments for relevant counterfactual questions. In our view, this provides a reminder that network models are not merely statistical hypothesis-testing tools, but flexible and powerful generative devices that can reveal emergent effects of multiple mechanisms on outcomes of interest in complex social systems.

In closing, we note that while statistical network models have seen great advances over the past 20 years, important challenges remain.  Among these is the problem of accounting for measurement error (a persistent challenge for the field since the famous call-to-arms of \cite{bernard1984problem}).  As with the vast majority of work in both social network analysis and demography, this paper considers the data as a fixed input without accounting for measurement error.  However, even Census data is imperfectly measured, a concern that becomes greater when considering the $\mathcal{O}(3000^2)$ migration rates that must be estimated to measure the U.S. county-level migration system. Assessing the nature and consequences of measurement error in migration networks remains an open problem, as does the estimation of count-valued ERGMs in the presence of measurement error. These would seem to be important directions for further work.

\hl{Likewise, in defining a network, one's choice of nodes and edges imposes a certain level of granularity on one's representation, which in turn impacts what effects it can distinguish \citep{butts:s:2009}.  Here, we examine the network of migration flows among U.S. counties, which could itself be seen as an aggregation of an ensemble of migrant flow networks for smaller subsets of the U.S. population; although we can hypothesize how these subflows contribute to the aggregate flow network, we are limited in our ability to disaggregate them here.  For example, we do not have information about whether and to what extent the larger migration flow from low-White-concentration counties to higher-White-concentration counties is driven by movement of the non-Hispanic White population, versus members of minority populations following on the heels of earlier migration by non-Hispanic Whites (an effect seen in some past research, e.g. \citet{woldoff_white_2011}). Distinguishing the migration patterns of different population groups within a joint model imposes significant challenges both from a data availability/accuracy and computational standpoint, but could provide further insights if feasible.
}

Last but not least, we note that there exist other states whose population redistribution patterns are stronger than California, such as New York State and Alaska, despite receiving less public attention. The impacts of the pandemic on internal migration, over both the short term and the long term (e.g., potential enhancement of ex-urban migration), are also critical research topics.  We encourage future research to examine these cases to offer a more comprehensive understanding regarding the evolution of the U.S. migration system and its implications for American society.

\singlespacing
\bibliographystyle{apa}
\bibliography{zotero}

\newpage
\section*{Appendix}

\begin{table}[h!]
\renewcommand{\thetable}{A\arabic{table}}  %
\setcounter{table}{0}  %
    \centering
        \caption{Valued ERGM for Inter-County Migration Flows, 2011-2015 (Full Model)}
    \label{tab:ergm.full}
    \begin{tabular}{lr@{\hskip -1pt}lr}
    \hline \hline
    & \multicolumn{2}{c}{Estimate}  & \multicolumn{1}{c}{Std Err} \\ \hline
    \emph{\textbf{Key Covariates in Table 2}} & \multicolumn{3}{c}{(included)}\\
    \emph{\textbf{Dependence Structures}} &&&\\
Mutuality & .047 & *** & .002 \\ 
  Waypoint flow & -.013 & *** & .001 \\ 
  Log(past migrant flow) & .300 & *** & $<$.001 \\  
  \emph{\textbf{Demographic Covariates}} &&&\\
  Origin log(population size) & .353 & *** & .002 \\ 
  Origin log(population size) & .374 & *** & .002 \\ 
  Destination log(population density) & -.081 & *** & .001 \\ 
  Origin log(population density) & -.055 & *** & .001 \\ 
  Destination PSR & .017 & *** & .001 \\ 
  Origin PSR & .017 & *** & .001 \\ 
  Destination log(immigrant inflow) & .057 & *** & .001 \\ 
  Origin log(immigrant inflow) & .043 & *** & .001 \\
  \emph{\textbf{Economic Covariates}} &&&\\
  Destination P(higher education) & .386 & *** & .012 \\ 
  Origin P(higher education) & .314 & *** & .013 \\ 
  Destination P(renter) & .404 & *** & .012 \\ 
  Origin P(renter) & .421 & *** & .012 \\ 
  Difference P(unemployment) & -1.229 & *** & .041 \\ 
  Origin P(unemployment) & -2.948 & *** & .052 \\ 
  \emph{\textbf{Geographical Covariates}} &&&\\
  Log(distance) & -.569 & *** & .001 \\ 
  Same state & .500 & *** & .002 \\ 
  Northeast & \multicolumn{3}{r}{(reference group)}\\
  Destination South & .257 & *** & .003 \\ 
  Origin South & .053 & *** & .003 \\ 
  Destination West & .384 & *** & .004 \\ 
  Origin West & .225 & *** & .004 \\ 
  Dstimation Midwest & .202 & *** & .003 \\ 
  Origin Midwest & .096 & *** & .003 \\
  \emph{\textbf{Baseline}} &&&\\
  sum & -1.421 & *** & .042 \\ 
  nonzero & -13.965 & *** & .028 \\ 
     \hline \hline
   \multicolumn{4}{l}{\emph{Note:} **$p<0.01$; ***$p<0.001$ (two-tailed tests).}  
    \end{tabular}
\end{table}

\end{document}